\documentclass[a4paper,11pt]{article}
\pdfoutput=1 % if your are submitting a pdflatex (i.e. if you have
\usepackage{jinstpub} % for details on the use of the package, please
\usepackage[dvipsnames]{xcolor}
\usepackage{upgreek}
\usepackage{gensymb}

\makeatletter
\define@key{Gin}{myresolution}{\pdfimageresolution=#1\relax}
\makeatother

\title{Characterization of 128x128 MM-PAD-2.1 ASIC: A Fast Framing Hard X-Ray Detector with High Dynamic Range}

\author[a,b]{D. Gadkari}
\author[c]{K.S. Shanks,}
\author[a,e]{H. Hu}
\author[a]{H.T. Philipp,}
\author[a]{M.W. Tate}
\author[b,1]{J. Thom-Levy,\note{Corresponding author.}}
\author[a,c,d]{and S.M. Gruner}
\affiliation[a]{Laboratory of Atomic and Solid State Physics, Cornell University,\\Ithaca, NY 14853, U.S.A}
\affiliation[b]{Laboratory for Elementary-Particle Physics, Cornell University,\\Ithaca, NY 14853, U.S.A}
\affiliation[c]{Cornell High Energy Synchrotron Source (CHESS), Cornell University,\\Ithaca, NY 14853, U.S.A}
\affiliation[d]{Kavli Institute at Cornell for Nanoscale Science, Cornell University,\\Ithaca, NY 14853, U.S.A}
\affiliation[e]{University of Chicago,\\Chicago, IL, U.S.A}

\emailAdd{jt297@cornell.edu}

\abstract{We characterize a new x-ray Mixed-Mode Pixel Array Detector (MM-PAD-2.1) Application Specific Integrated Circuit (ASIC). Using an integrating pixel front-end with dynamic charge removal architecture, the MM-PAD-2.1 ASIC extends the maximum measurable x-ray signal (in 20 keV photon units) to > 10$^{7}$ x-rays/pixel/frame while maintaining a low read noise across the full dynamic range, all while imaging continuously at a frame rate of up to 10 kHz. The in-pixel dynamic charge removal mechanism prevents saturation of the input amplifier and proceeds in parallel with signal integration to achieve deadtime-less measurements with incident x-ray rates of > 10$^{10}$ x-rays/pixel/s. The ASIC format consists of 128$\times$128 square pixels each 150 $\upmu$m on a side and is designed to be 3-side buttable so large arrays can be effectively tiled. Here we use both laboratory x-ray sources and the Cornell High Energy Synchrotron Source (CHESS) to characterize two single ASIC prototype detectors for both low (single x-ray) and high incident flux detection. In the first detector the ASIC was solder bump-bonded to a 500 $\upmu$m thick Si sensor for efficient detection of x-rays below ~20 keV, whereas the second detector used a 750 $\upmu$m thick CdTe sensor for x-rays above $\sim$ 20 keV. }


\begin{document}
\maketitle
\flushbottom

\section{Introduction}
\label{sec:intro}

Synchrotron x-ray facilities are moving towards higher x-ray energies, higher repetition rates and increased brilliance \cite{shin2021}. Accordingly, we seek to develop x-ray area detectors with wide dynamic range, high stopping power across a wide range of x-ray energies, high frame rates, and accurate measurement of signals ranging from single x-rays per frame to high continuous and instantaneous hit rates.

In a hybrid direct x-ray pixel array detector (PAD), a sensor layer (e.g., of Si or CdTe) absorbs x-rays to produce electron-hole pairs in numbers proportional to the stopped x-ray energy. An array of pixelated contacts on the backside of the monolithic sensor layer is electrically connected via metal bump-bonds to a corresponding array of pixels in an Application Specific Integrated Circuit (ASIC) layer. Each ASIC pixel has its own circuitry (figure \ref{fig:asic}) which processes the charge collected from the sensor pixel, with all pixels operating in parallel during the x-ray exposure. The bonded sensor-ASIC combination unit, or ``hybrid module'' is the fundamental PAD unit. The ASIC die is limited in size by the integrated circuit fabrication process to typically one or two cm across. However, large area detectors may be made by side-by-side tiling of hybrid modules. The ultimate performance of a given detector is limited by the characteristics of the sensor and ASIC that comprise the hybrid module. 

Broadly speaking, PADs fall into two groups: those that count, one-by-one, the current pulses resulting from each x-ray (``photon counters'') and those that integrate the current for a period of time and then digitize the result (``integrators''). Both photon counters and integrators can have good single x-ray detection capability with the appropriate choices in the front-end amplifier. 

The photon counting detectors that predominate at storage ring sources can accurately measure flux rates of $\sim$ 10$^{6}$-10$^{7}$ photons/pixel/second \cite{ballabriga2016} before x-ray processing times compromise the measurement and introduce counting non-linearities. These flux rates are insufficient for signals encountered at x-ray free electron lasers (XFELs) and are often insufficient at third generation and upgraded storage rings.  This has motivated us to focus our effort on developing integrating detectors that do not have the limitations of photon counting detectors.

A first generation Mixed Mode Pixel Array Detector (MM-PAD) platform has been previously developed and was initially bonded to Si sensors \cite{schuette2008,tate2013}. To extend the capabilities at higher x-ray energy (> 20 keV), CdTe sensors were subsequently bonded to the MMPAD \cite{becker2016}. Both types of sensors have demonstrated excellent performance in a wide variety of synchrotron radiation experiments \cite{Giewekemeyer2014,Giewekemeyer2019,chatterjee2017}. The MM-PAD-1 ASIC operates at frame-rates up to 1.1 kHz, although there is no capability to expose while reading out, limiting the duty cycle at high frame rates. Further, the in-pixel charge removal circuitry has a minimum operation period which limits the sustained signal rate to 4$\times$10$^{8}$ x-rays/pixel/s (in 8 keV x-ray units). Updates to the ASIC electronics have been explored over the years \cite{weiss2017,weiss2017II} with the aim of increasing the sustained photon rate, improving the duty cycle at high frame rates as well as increasing the overall frame rate. This paper is a report on the characterization of the MM-PAD-2.1 ASIC using Si and CdTe single-module detectors that were built for testing and prototyping purposes. These ASICs will be used to produce several tiled detectors to be installed at beamlines at CHESS and at the Advanced Photon Source (APS) at Argonne National Laboratory.

\section{Detector overview}
\label{sec:arch}
%----------------------Basic pixel design------------------------
\subsection{ASIC}
\label{subsec:asic}

\begin{figure}
\centering 
\includegraphics[width = 0.7\textwidth]{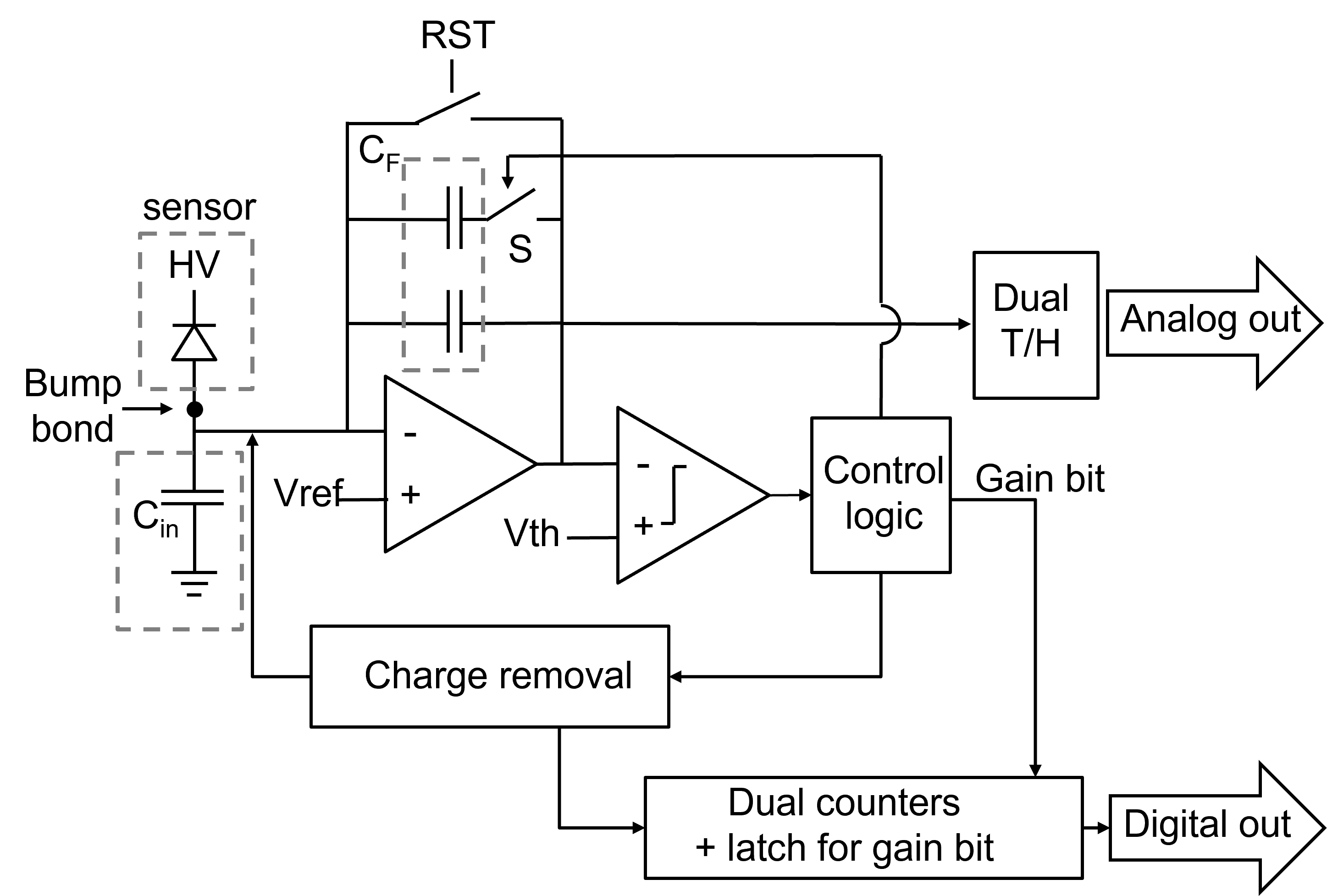}
\caption{\label{fig:asic} MM-PAD-2.1 single pixel circuit diagram. The sensor is represented by a diode on the pixel input node. The front-end has a charge integrating amplifier monitored by a voltage comparator. The high-gain and low-gain feedback capacitors are labeled collectively as C$_{F}$ for simplicity. The parasitic capacitance at the pixel input node, C$_{in}$, is shown for reference and has contributions from the sensor capacitance, CMOS circuitry and bond capacitance. When the voltage comparator is triggered for the first time, the adaptive gain switches to low-gain mode by closing CMOS switch S, setting the in-pixel gain-status bit. All subsequent comparator triggers increment the digital counter and initiate a charge removal ($\sim$ 8 MeV equivalent charge), with integration of charge from the sensor continuing uninterrupted. The dual in-pixel counters and track-and-hold circuit enable expose-during-readout operation.}
\end{figure}

The conceptual design of the MM-PAD-2.1, along with measurements using a small-scale 16$\times$16 pixel prototype, is described in \cite{shanks2021}. The block diagram of the MM-PAD-2.1 pixel is shown in figure \ref{fig:asic}. A comparison between the measured MM-PAD-1 and MM-PAD-2.1 systems is listed in table \ref{table:1}. The MM-PAD-2.1 uses a Class AB transimpedance amplifier with common mode feedback at the front-end. It has two-stage adaptive gain incorporated, similar to the approach of AGIPD \cite{agipd} and JUNGFRAU \cite{jungfrau}, with the nominal feedback capacitance being 40 fF and 880 fF in high and low gain modes, respectively. As the signal is integrated across the feedback capacitors, the front-end output is compared to a threshold voltage, V$_{th}$, which is set globally across the ASIC. When the output of a pixel crosses this threshold for the first time, that pixel switches into low gain mode and a pixel gain-status bit is set. At each subsequent crossing of the threshold, the charge removal circuit is triggered which removes a fixed amount of charge from the front end feedback capacitor. With each charge removal cycle, one of the two in-pixel 16-bit counters is incremented. The amount of charge removed per cycle is determined by external biases applied globally across the whole ASIC. Adaptive gain allows a larger amount of charge to be removed per cycle as compared to the MM-PAD-1. This, coupled with faster operation of the charge removal circuitry (100 MHz vs. 2 MHz for the MM-PAD-1), results in a pixel which is now capable of operating at photon fluxes of $>$ 10$^{10}$ 20 keV photons/pix/s. To allow exposure during readout, dual in-pixel counters for the digital data, and dual track-and-hold circuits for the residual analog data were implemented. The analog data stream is digitized by ADCs off-chip at a rate of 10 MHz. The readout of the array is broken into 16 banks of 8$\times$128 pixels each with one differential analog output and one LVDS digital output per bank. This architecture allows 10 kHz framing with the duty cycle for exposure at > 99 \%. In the present implementation, the detector is run with a minimum 1 $\upmu$s dead time between frames, allowing continuous operation with 99 \% duty cycle at 10 kHz frame rate. Additionally, logic was added to allow the pixels to function correctly with either the collection of holes or electrons from the sensor.

\begin{table}[h!]
\centering
\begin{tabular}{ c c c  } 
 \hline
  & MM-PAD-1 & MM-PAD-2.1 \\
  & (8 keV equivalent) & (20 keV equivalent) \\
 \hline
 Format & 128$\times$128 & 128$\times$128 \\ 
 Pixel pitch & 150 $\upmu$m  & 150 $\upmu$m \\ 
 Sensor & Si,CdTe  & Si,CdTe \\ 
 Charge Collection & holes only & electrons or holes \\ 
 Frame rate (FR) & 1.1 kHz & 10 kHz \\ 
 Duty Cycle, max FR & 0\% & 99\% \\ 
 Read noise & 0.16 photon & 0.13 photon \\ 
 Well capacity & 4.7$\times$10$^{7}$ photons & 2.2$\times$10$^{7}$ photons \\ 
 Instantaneous count rate & > 10$^{12}$ ph/pix/s & > 10$^{12}$ ph/pix/s \\ 
 Sustained count rate & 4$\times$10$^{8}$ ph/pix/s & > 3$\times$10$^{10}$ ph/pix/s \\ 
 \hline
\end{tabular}
\caption{MM-PAD-1 and MM-PAD-2.1 specification table. Values listed here are for 128$\times$128 pixel  ASICs bonded to a Si sensor. The values for the MM-PAD-1 are from \cite{tate2013}. ``Instantaneous count rate'' refers to the effective photon rate the detector is capable of measuring in a single pulse of $\leq$ 100 ps, i.e. in a single storage ring pulse (typically of order 10-100 ps) or XFEL pulse (typically $\leq$ 100 fs).}
\label{table:1}
\end{table}
 
\subsection{ASIC-sensor hybrids and test system}
\label{subsec:ss_det_sys}

The full-scale MM-PAD-2.1 will consist of a 2$\times$3 array of tiled modules (256 $\times$ 384 pixels), with a tile comprised of a 128$\times$128 pixel hybrid module consisting of an ASIC bonded to a CdTe sensor for high-energy x-ray (> 20 keV) applications. Here, we describe characterization measurements for single 128$\times$128 pixel hybrid modules consisting of an MM-PAD-2.1 ASIC bonded to sensors either of Si (500 $\upmu$m thick, hole-collecting, fabricated by SINTEF, Oslo, Norway and solder bump-bonded by Micross, North Carolina, USA) or In Schottky CdTe (750  $\upmu$m thick, hole-collecting, fabricated by Acrorad, Okinawa, Japan and solder-bump bonded by Direct Conversion, Espoo, Finland). Each of the ASIC-sensor hybrid modules is attached to an aluminum heat sink using thermally conductive silicone adhesive and wire bonded to a printed circuit board. A thermoelectric cooler is used to keep the Si sensor at -20$^{\circ}$C and the CdTe sensor at 0$^{\circ}$C. The sensors are over-depleted by applying a voltage of 150 V across the Si and 400 V across the CdTe sensor. The temperature and the bias across the CdTe sensor were determined in previous studies \cite{becker2016} to minimize polarization effects. The printed circuit board is mated with a secondary long circuit board that is used for routing electric signals to and from the detector module. A dedicated FPGA is used to assemble and organize the data stream, which is then sent via a Cameralink connection to a PC for storage and analysis.

\section{ASIC characterization}
\subsection{Signal measurement and analysis techniques}
\label{subsec:ana_tech}
As described in section \ref{subsec:asic}, the MM-PAD-2.1 output consists of two data streams for each pixel. The first is the residual analog signal which is digitized off-chip to 14 bits and is reported in analog-to-digital units (ADU). The second stream is the count of the number of charge removals combined with the gain bit of the integration stage. These streams are scaled together to a number proportional to the total deposited energy per pixel as described in section \ref{subsec:calib_ana_dig}. An average of $\sim$ 100 dark images is subtracted from the combined  data streams. This subtraction removes any accumulated dark current as well as the zero point offset of the circuitry. A frame-by-frame common-mode offset ($<$ 10 ADU) is observed even after background subtraction. This can be corrected using a statistical ``debouncing'' technique for frames having a sufficient number of pixels with no recorded photons which allows an adjusted ``zero'' level to be determined. The magnitude of the common mode offset is small enough to be ignored for those frames with a higher level of illumination.

\subsection{Pixel gain and noise measurements}
\label{subsec:gain}
Pixel gain was determined in high-gain mode via x-ray spectral histograms. For the Si sensor system, the gain measurement was first done for a subset of pixels. A 150 $\upmu$m thick tungsten pinhole mask with 25 $\upmu$m diameter hole size was placed in front of the detector. To reduce effects of charge sharing between pixels, data analysis was limited to those spots which were isolated to a single pixel. These spots were produced with monochromatic 22.16 keV x-rays coming from a graphite monochromator illuminated by a 50 W silver anode tube operated at 45 kV and 0.4 mA. The resulting exposures are governed by a Poisson distribution in the number of x-rays per frame, with pixels recording no x-rays in most frames, 1 x-ray in some frames, 2 x-rays in even fewer frames etc. Ten thousand x-ray frames along with 200 dark (no signal) frames were captured. 

\begin{figure}
\centering 
\includegraphics[myresolution = 300]{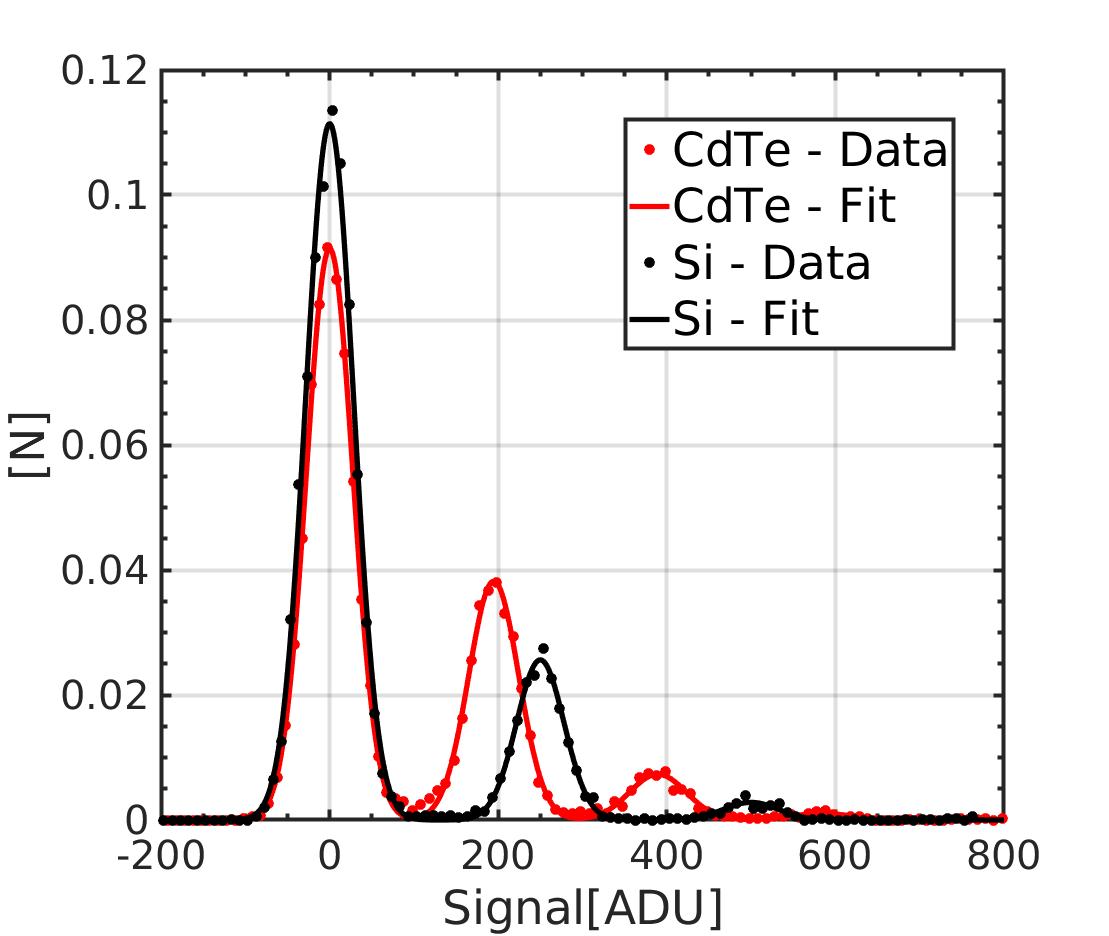}
\caption{\label{fig:pinhole_spectra} Pinhole spectra using 22.16 keV Ag K-$\alpha$ x-rays for pixels in the (black) Si sensor system, and (red) CdTe sensor system. The peaks at 0 ADU correspond to frames with no incident x-ray. The peaks at 248 ADU (black) and 388 ADU (red) correspond to frames with one x-ray photon. The average gain for the pixels in high-gain mode is measured to be 11.2 ADU$_{H}$/keV for the Si sensor system and 8.32 ADU$_{H}$/keV for the CdTe sensor system.}
\end{figure}

From the set of background subtracted and debounced frames, a histogram of the recorded signal in each pixel is produced (in units of ADU$_{H}$, where ADU$_{H}$ refers to analog-to-digital units in high-gain). Each histogram consists of a discrete series of peaks as seen in figure \ref{fig:pinhole_spectra}, where the position of the peaks on the x-axis corresponds to the number of the absorbed photons. Hence, the first peak centered at 0 ADU$_{H}$, corresponds to 0 absorbed photons, the next peak to 1 absorbed photon and so on. The spacing between any pair of consecutive peaks then corresponds to the signal recorded by the pixel in ADU$_{H}$ for a single absorbed photon and is the pixel gain [ADU$_{H}$/keV]. A sum of Gaussians with relative intensities following a Poisson distribution is fit to the data to extract the gain, as described previously in \cite{gadkari2020}. Figure \ref{fig:pinhole_spectra} (black) shows the photon spectra of one of the Si sensor pixels obtained using the tungsten mask. 

To gather gain information from all the pixels in the Si sensor array, a second set of measurements was conducted. Here, an approximately uniform illumination was made using a 50 W molybdenum anode x-ray tube operated at 25 kV and 0.365 mA.  A 200 $\upmu$m thick zirconium filter was placed downstream of the x-ray tube to improve the fraction of x-rays from the 17.48 keV Mo K$\alpha$ line. Frames were taken such that there was less than 0.1 x-ray/pixel/frame to minimize overlap of x-ray events. The histogram of pixel values contains x-ray events which share charge across pixel boundaries, but a large fraction of events are fully contained within a single pixel, resulting in a sharp photon peaks. The gain measured by isolated illumination was found to be within 1\% of the gain measured by flood illumination. The average pixel gain across the 128$\times$128 pixel detector was measured to be 11.2 $\pm$ 0.1 ADU$_{H}$/keV for the Si sensor. 

For the CdTe sensor system, pixel gain was measured using a 150 $\upmu$m thick tungsten mask with 75 $\upmu$m diameter hole size to isolate the pixels. The central pixel regions were illuminated with 22.16 keV Ag K$\alpha$ line, with the silver tube operating at 45 kV and 0.4 mA, and the K$\alpha$ selected using the graphite monochromator. The frame integration time was set to 10 ms and 10,000 x-rays frames, along with 500 dark frames, were captured for the measurement. Figure \ref{fig:pinhole_spectra} (red) shows the photon spectra of one of the CdTe sensor pixels. The average gain measured for the CdTe system is 8.32 $\pm$ 0.02 ADU$_{H}$/keV. The measured ratio of 0.74 $\pm$ 0.01 between the gains of CdTe and Si is slightly less than the expected value of 0.817, which is the ratio between the pair creation energies of CdTe and Si \cite{klein1968}. This could be due to a decrease in charge collection efficiency (CCE) in the CdTe material due to the presence of traps or also possibly due to higher input capacitance C$_{in}$ in the CdTe sensor, resulting in a reduced CCE \cite{gadkari2020}.

Read noise was measured for both the Si and CdTe sensor systems using 1000 dark frames for each type. The frame integration time was set to 100 $\upmu$s so that the noise contribution due to dark current was negligible. The average noise across the Si sensor system was measured to be 2.5 $\pm$ 0.07 keV. The average noise in the CdTe sensor system was 3.72 $\pm$ 0.07 keV. Thus, the Si and CdTe systems provide an SNR of $\geq$ 2 for x-ray energies above 5 keV and 8 keV respectively.

\subsection{Analog and digital data merging}
\label{subsec:calib_ana_dig}
 
As described in section \ref{subsec:asic}, the MM-PAD-2.1 output has analog and digital data streams which need to be scaled and added together to get the total incident signal. The use of an adaptive gain stage requires another scaling factor between the analog data obtained in low or high gain mode. As a matter of convenience, we scale both the low-gain analog data and the digital data to data taken in the high-gain mode, expressed in units of ADU$_{H}$. The total pixel signal is found using the following formula :

\begin{equation}
\label{eq:scale_data}
S_{total} = S_{ana}(1-G)+(C_{A}S_{ana}+O_{A})G+C_{D}S_{dig},
\end{equation} 
where $S_{total}$ is the total signal incident on the pixels (in units of ADU$_{H}$), $S_{ana}$ is the residual analog output from the pixel (digitized via off-chip ADC) corresponding to the integrated charge remaining in the front-end at the end of an integration cycle after any and all charge removal steps, $S_{dig}$ is the number of digital counts (number of charge removals in the pixel), $G$ is the gain bit where 0 indicates high-gain mode and 1 indicates low-gain mode, $C_{A}$ is a scale factor between the analog data in high and low gain modes, $O_{A}$ is the offset between high gain and low gain analog data, and $C_{D}$ is a calibration factor to scale digital data with the high-gain analog data. The calibration factors provide a linearized output for each pixel. An additional constant per pixel is needed to normalize the response of each pixel relative to all other pixels in the array. The inter-pixel calibration is taken care of by multiplying $S_{total}$ with the pixel gain, scaling the total signal from ADU$_{H}$ to keV, as described in section \ref{subsec:gain}.

To determine $C_{A}$, $O_{A}$ and $C_{D}$ for each pixel, a data set is taken consisting of flood field illuminations for a series of integration times. For the Si sensor system, the detector was illuminated by a flood field from a 50 W molybdenum anode tube operating at 25 kV and 0.210 mA. The frame integration time was varied from 100 $\upmu$s to 9 s with 200 x-ray frames captured for each integration time. Data times were chosen to span each of the operating regions with multiple integration time (the regions to be covered are the high gain analog region, the low gain analog region, and the digital counting region). For the CdTe sensor system, the detector was exposed to a flood field from a 50 W silver anode tube, operated at 30 kV and 0.2 mA. The integration time was varied from 100 $\upmu$s to 100 ms for this system, with 200 x-ray frames captured for each integration time. 

\begin{figure}
\centering 
\includegraphics[width=\textwidth]{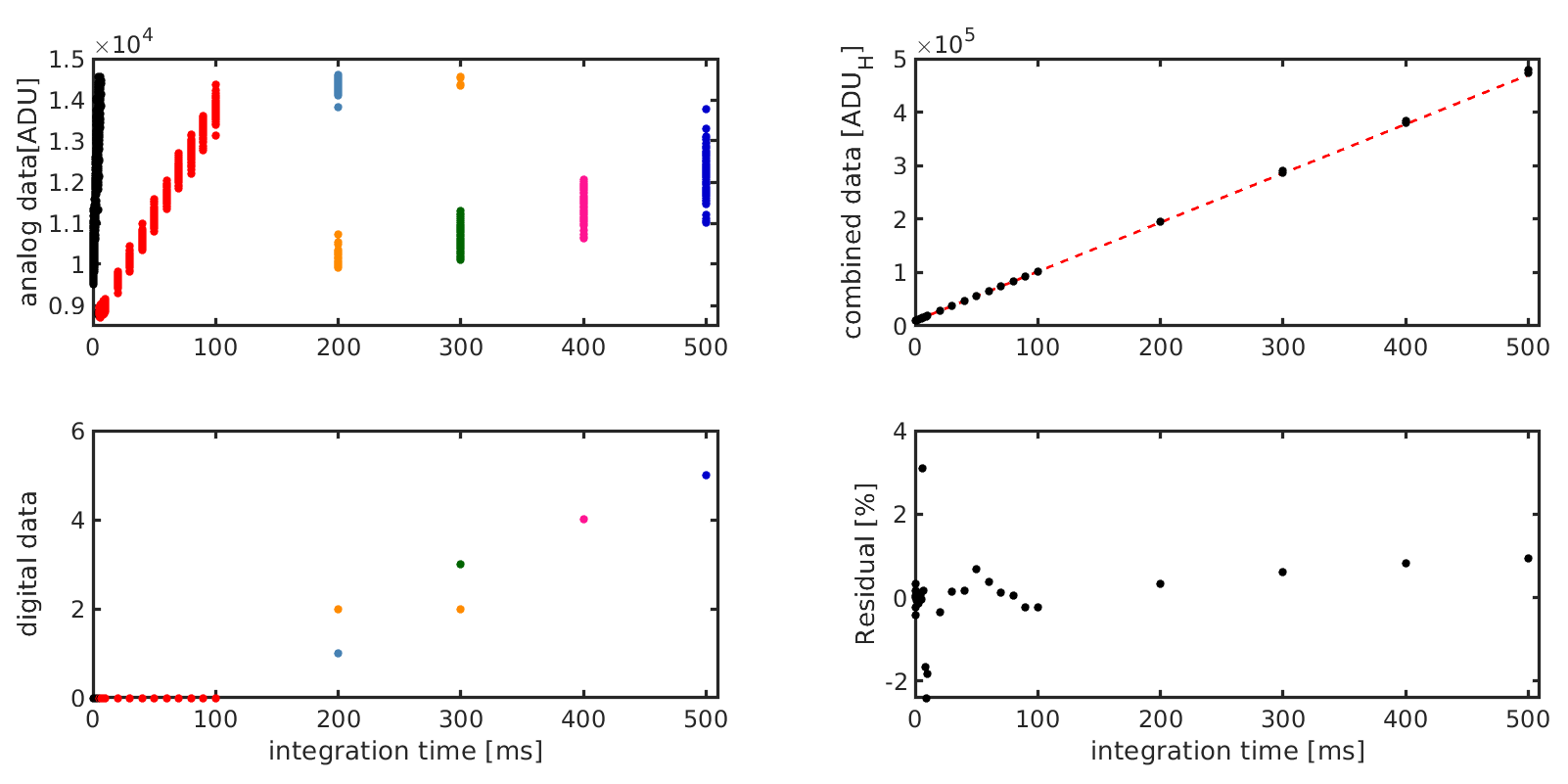}
\caption{\label{fig:cali_signal} Pixel response as a function of integration time for constant illumination. Data from a single pixel in the Si sensor system. (Top left) Unscaled raw analog data (in units of raw ADU, i.e. the direct raw output from the ADC) as a function of integration time. (Bottom left) Digital data from this pixel as a function of integration time. Black points in both plots correspond to all frames where the pixel was in high gain and red points correspond to all frames where the pixel was in low gain but did not have any charge removals. All the subsequent points are color coded and are matched between the analog and digital data according to the number of charge removals which occurred in the frame. (Top right) Total signal intensity, averaged across all frames for a given integration time, obtained by combining and scaling the analog and digital data together to ADU$_{H}$. A straight line (red dashed line), fit only to the high gain region, is extrapolated for all integration times. (Lower right) Residual between the fit and the combined data, shown as percentage of the fit value.}
\end{figure}

Figure \ref{fig:cali_signal} shows the residual analog data, $S_{ana}$ (top left), and the digital data, $S_{dig}$ (bottom left), for different frame integration times for a typical pixel in the Si system. The data is color coded according to gain bit and digital count for a given frame. The spread in the data points at any given time corresponds to the shot noise in the x-ray signal per frame. The analog data shows a steep rise at short times (black points) with the system in high gain. Red points show the response after the system has transitioned into low gain. $C_{A}$ was determined by taking the ratio of the slopes in these two regions. $O_{A}$ is the offset needed to align the scaled low-gain data with the high-gain curve. Finally, $C_{D}$ was determined by a least squares fit of the analog data (scaled to high-gain units) and the digital data. The average $C_{A}$, $O_{A}$ and $C_{D}$ across the detector for the Si sensor system were determined to be 17.82 $\pm$ 1.04 ADU$_{H}$/ADU$_{L}$, (-1.4 $\pm$ 0.1)$\times$10$^{4}$ ADU$_{H}$ and (7.5 $\pm$ 0.1)$\times$10$^{5}$ ADU$_{H}$ (i.e. 6.7 MeV charge per removal). For the CdTe sensor system $C_{A}$, $O_{A}$ and $C_{D}$ were determined to be 18.98 $\pm$ 2.51 ADU$_{H}$/ADU$_{L}$, (-1.5 $\pm$ 0.2 )$\times$10$^{5}$ ADU$_{H}$, and (6.7 $\pm$ 0.1)$\times$10$^{5}$ ADU$_{H}$ (i.e. 8.1 MeV charge per removal). The variations reported in the calibration constants are due to pixel-to-pixel nonuniformities. The measured constants have been observed to be stable over the duration of the measurements and the uncertainty in the determination of any given constant is less than 1\%.

 %MWT - use previous sentence, not the following - \textcolor{red}{The calibration constants obtained are consistent with the once measured for the small-scale and have been consistent and stable across all the other measurements done with this detector}.%

$C_{A}$ can also be used to get an estimate of the pixel input capacitance, C$_{in}$. C$_{in}$ (labeled in figure \ref{fig:asic}) is the parasitic capacitance at the pixel input node and has contributions from the CMOS circuitry, the sensor bulk, the bump-bonds and the bond pads as discussed in \cite{gadkari2020}. C$_{in}$ causes a decrease in the charge collection efficiency as the photogenerated charge gets divided between the feedback capacitor C$_{F}$ and C$_{in}$. This system can be modeled as a circuit with zero C$_{in}$ and an effective feedback capacitor C$_{Feff}$, given by :

\begin{equation}
\label{eq:cfeff}
C_{Feff} = C_F\frac{C_{in}+(1+A)C_F}{(1+A)C_F},
\end{equation} 
where $A$ is the open-loop voltage gain. For the MM-PAD-2.1, the feedback capacitance in high-gain mode, C$_{FH}$, is 40 fF and in low-gain mode, C$_{FL}$, is 880 fF.  From standard SPICE simulations of the amplifier, $A$ is expected to be from 34 to 39 dB over the range of mobility parameters for the process used in the fabrication of the ASIC. We assume that the effective feedback capacitance in low-gain mode, C$_{FLeff}$, is 880 fF, its as-designed value, which is then scaled by $C_{A}$ to obtain C$_{FHeff}$, the effective feedback capacitance in high-gain mode. Equation \ref{eq:cfeff} is then used to estimate C$_{in}$. For the Si sensor system, C$_{FHeff}$ is estimated to be 49.38 $\pm$ 2.88 fF, yielding a C$_{in}$ estimate of $\sim$240-450 fF. For the CdTe system, C$_{FHeff}$ is estimated to be 46.36 $\pm$ 6.13 fF and there is a large uncertainty in the estimated C$_{in}$ ($\sim$10-510 fF) for the system.

\subsection{Edge response}
\label{subsec:resol}
\begin{figure}
\centering 
\includegraphics[width=\textwidth]{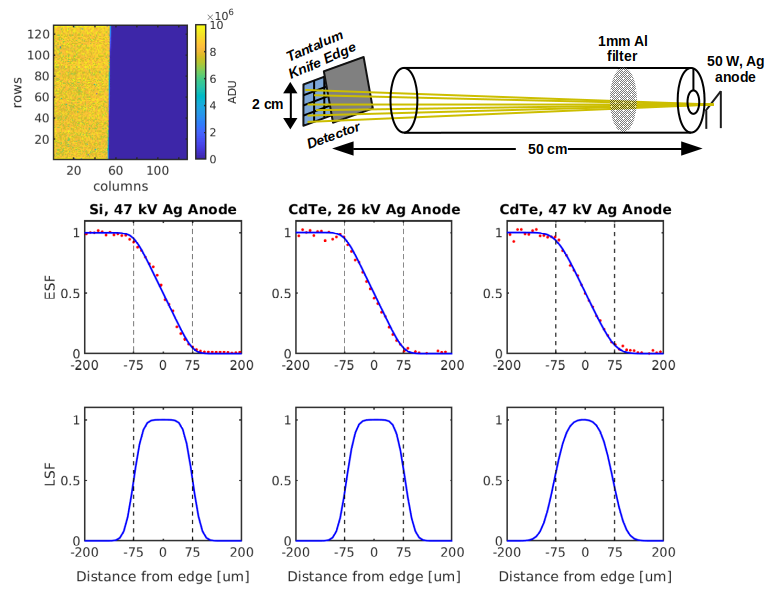} %resolution_Si_CdTe2.png
\caption{\label{fig:edge_photo} Detector resolution measurements. (Top right) Edge response measurement setup. An x-ray tube, with optional Al filter, flood illuminates the detector. A tantalum knife is placed in front of the detector at a small angle relative to the detector pixel columns. (Top left) An image of the detector response for the Si system. (Middle row) The edge spread function for the Si (left) and CdTe (middle at 26 kV tube bias without filter and right at 47 kV tube bias with 1 mm Al filter). The dashed lines represent the boundaries of the 150 $\upmu$m pixel. Also shown is the best fit of a convolution of an ideal pixel response (a one pixel wide ramp function) with a Gaussian, to the data. (Bottom row) The respective line spread functions obtained by taking the derivative of each of the functions fit to the ESF measurements.}
\end{figure}

The spatial resolution of the detector was determined by measuring the edge spread function (ESF), which is the response to a step function of x-ray intensity vs. position. The ESF of a detector is used to determine the effective charge cloud size in the sensor for the corresponding incident x-ray energies, which has a direct impact on the spatial performance of the sensor. To measure the ESF, a tantalum knife edge was placed in front of the detector, slightly tilted ($\theta$=0.94$^{\circ}$ for the Si sensor, $\theta$=1.36$^{\circ}$ for the CdTe sensor) with respect to the columns such that a few columns were partially covered by the knife edge. The particular angle of tilt is not important as long as it is small. Angles around 1-2$^{\circ}$ allow the edge spread function to be sampled at sub-pixel resolution in a single image as pixels along the knife edge are covered fractionally more and more by the knife as one translates along the knife edge. The Si detector was illuminated by a 50 W silver anode tube, positioned 50 cm away from the detector and operated at 47 kV and 0.4 mA. The CdTe detector was illuminated by a 50 W silver anode, positioned 50 cm away from the detector, and operated at two different tube voltages of 26 kV and 47 kV, at 0.4 mA. An additional filter of 1 mm aluminum was used at the 47 kV setting for the CdTe, to increase the average energy of the incoming beam.  For Si, the frame integration time was set to 10 s. 1000 x-ray frames and 100 dark frames were captured, while for CdTe, the frame integration time was set to 20 ms, and 200 x-ray frames and 100 dark frames were captured. Figure \ref{fig:edge_photo} (top left) shows the total signal in the Si detector when partially covered by the knife edge. The knife-edge data is normalized by dividing by a data set taken without the knife.

The distance from the center of each pixel in the image to the knife-edge (along the normal to the edge) can be computed. Data within 2 pixels of the edge is shown in figure \ref{fig:edge_photo} (middle). The ideal pixel response would be a one pixel wide linear ramp. Charge spread within the sensor can be seen with the deviations from this response. The convolution of the ideal response ramp with a Gaussian with varying width was fit to the ESF data. The width of the fitted Gaussian is a measure of the charge spread in the sensor. The fitted functions had a width of 20 $\upmu$m for Si, 20 $\upmu$m for CdTe at 26 kV tube bias, and 30 $\upmu$m for CdTe at 47 kV tube bias. The reduced resolution in the last case is expected since a significant fraction of the incident x-rays are above the energy threshold for fluorescence production in Cd or Te. These fluorescent x-rays spread within the sensor material and are reabsorbed a distance away from the initial interaction point. 

One can determine the line spread function (LSF), which is the response to an infinitely thin line illumination,  by taking the derivative of the fit to the ESF. This is shown in figure \ref{fig:edge_photo} (bottom). Again, one can observe the reduced resolution for high energy x-rays in the CdTe system.

\subsection{Fast-framing characterization}
\begin{figure}
\centering 
\includegraphics[myresolution = 200]{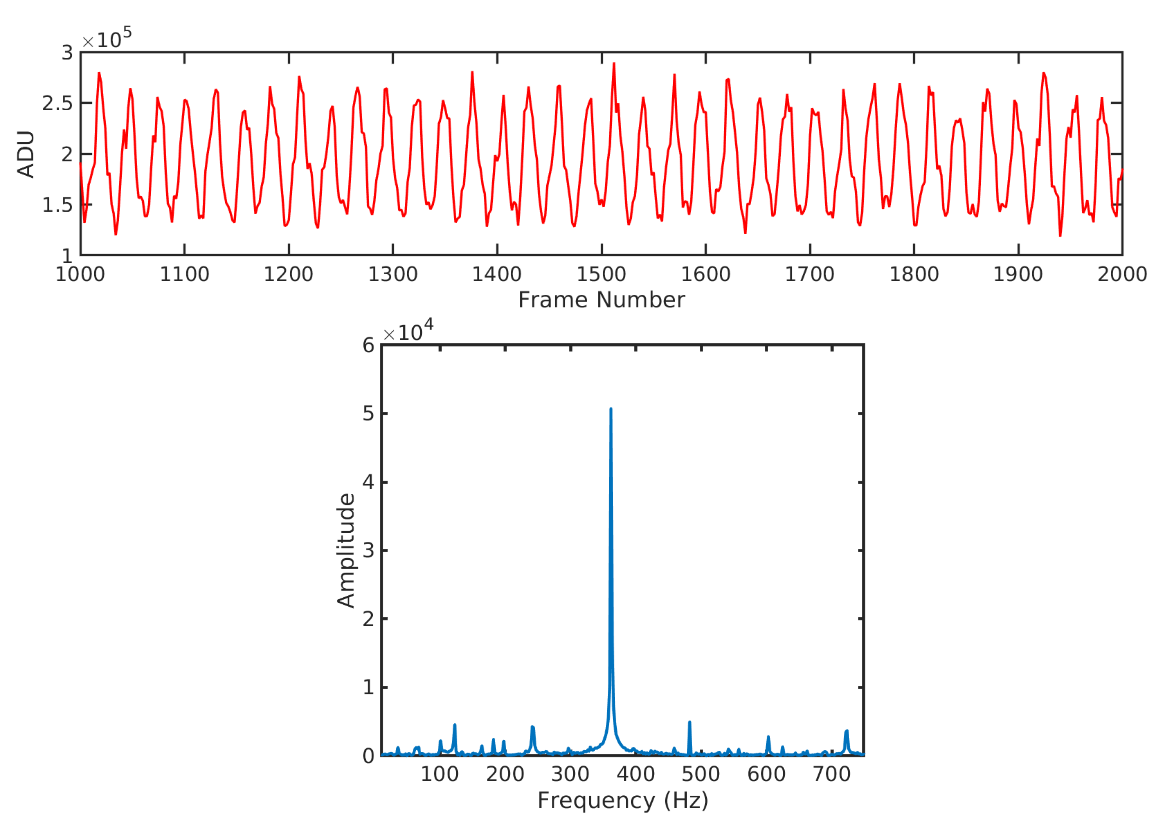}
\caption{\label{fig:rotanode} (Top) Background subtracted signal intensity per frame, as recorded by the Si MM-PAD-2.1 vs frame number, with a frame integration time of 100 $\upmu$s. The copper rotating anode x-ray source intensity varies by 360 Hz, as also recorded by the detector. (Bottom) A fast Fourier transformation of the signal above extracts a signal frequency of 361$\pm$1 Hz, which matches very closely to the source intensity variation. }
\end{figure}

The fast-framing capabilities of the Si MM-PAD-2.1 ASIC system were shown using an in-house Rigaku RU-3HR copper rotating anode x-ray source. This x-ray source has a 3-phase power supply input which results in an output source intensity variation of 360 Hz. The timing performance of the MM-PAD-2.1 detector was measured by extracting this frequency of source intensity variation. The x-ray tube was operated at 38 kV and 40 mA. The unfocused beam spanned 12$\times$8 pixels (FWHM), with an average incident flux of $\sim$10$^{7}$ ph/s/mm$^{2}$. Figure \ref{fig:rotanode} (top) shows the background subtracted signal summed across 20$\times$14 pixels illuminated by the beam. The signal was observed to vary periodically over time, within $\sim \pm$35$\%$ of the average incident signal. 5000 frames were captured, with a measured frame rate of 9.9 kHz, and a frame integration time of 100 $\upmu$s. The fast Fourier transform (FFT) algorithm from Matlab was then used to extract the frequency of the signal variation, shown in figure \ref{fig:rotanode} (bottom). The leading frequency extracted from the signal variation was 361$\pm$1 Hz, close to the expected 360 Hz. No other notable frequency peaks were extracted from the FFT, and the other frequencies extracted by it can be attributed to noise and pickup in the system.

\subsection{Performance of CdTe MM-PAD-2.1 for high-energy x-rays}
The performance of CdTe MM-PAD-2.1 was measured at the Cornell High Energy Synchrotron Source (CHESS) for high-energy x-rays under both high-flux (up to 10$^{12}$ keV/s/mm$^{2}$) and low-flux (average of $\ll$1 x-rays/pixel/100 $\upmu$s) illumination at beamline 3A. The beamline is fed with a 1.5m long CHESS Compact Undulator (CCU) \cite{Temnykh2015,Temnykh2013} and a silicon <220> monochromator is used to isolate 61.332 keV x-rays. The incoming beam was not perfectly monochromatic as $\sim$2.5\% of the total photons in the beam were 122.664 keV second harmonic x-rays. Guard slits were used to set the beam size 1 m upstream of the detector, and steel attenuators of different thickness were used to vary the incident flux. An ionization chamber, immediately downstream of the slits was used to measure the incident flux for the high-flux characterization measurements. 

\subsubsection{Low-flux performance}
\begin{figure}
\centering 
\includegraphics[width = \textwidth]{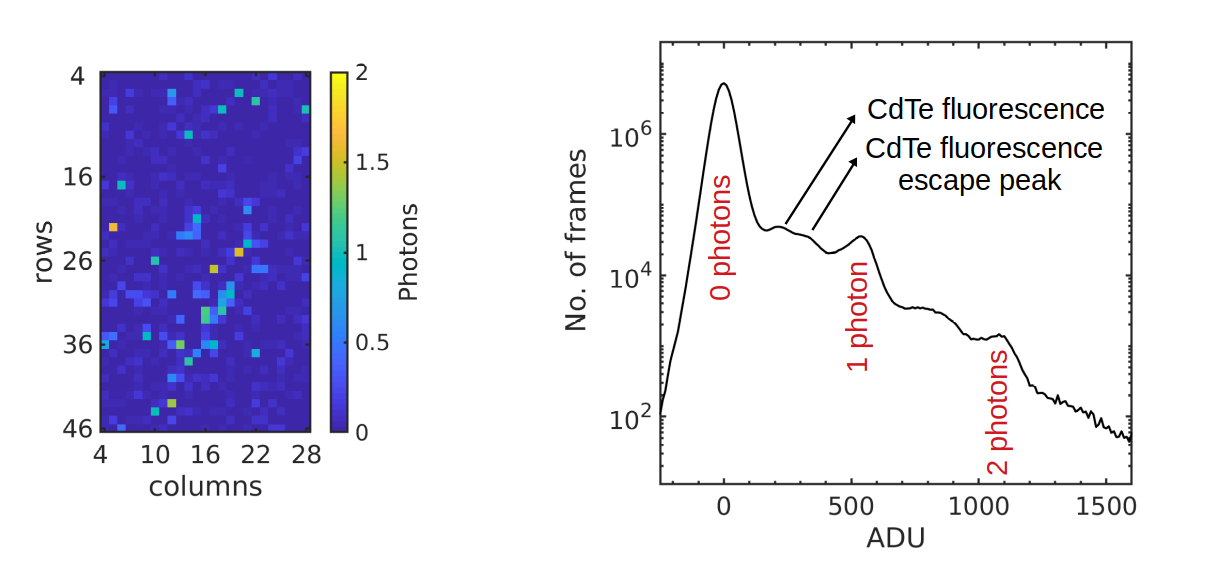}
\caption{\label{fig:CHESSspectra}(Left) An example of a single low-flux signal frame in a CdTe MM-PAD-2.1, shown for a subset of pixels, with an incident x-ray energy of 61.332 keV and a 1.4 mm thick aluminum attenuator attached in front of the detector. The z-axis is depicted in 61.332 keV photon equivalent signal, and the bright spots in the image correspond to single or double x-ray photon events. (Right) Histogram of the recorded signal in a subset of pixels (left image) over 40,000 frames showing not only the 0,1,2 photon peaks, but also the CdTe fluorescence and escape peaks, as labeled.}
\end{figure}

For low-flux characterization, the frame integration time was set to 100 $\upmu$s, with 100 $\upmu$m $\times$ 100 $\upmu$m beam size, no steel attenuator and a 1.4 mm thick aluminum attenuator attached immediately upstream of the detector. 40,000 frames were captured, with an average incident flux of $\ll$1 x-rays/frame for 99\% of the pixels and $\sim$2 x-rays/frame for the rest of the pixels. Figure \ref{fig:CHESSspectra} (left) shows an example of a single signal frame in the detector, where the signal is depicted in 61.322 keV photons equivalent. The spectrum of the recorded signal from the incident 61.332 keV x-rays across the 40,000 frames is shown in figure \ref{fig:CHESSspectra} (right) for a subset of pixels (43$\times$25 pixels).  Note that the histogram is comprised primarily from pixels with a flux of $\ll$1 x-rays/pixel/frame. CdTe exhibits fluorescence at the K$\alpha$ energies of Cd at 23.2 keV and Te at 27.5 keV. Apart from the 0, 1 and 2 photon peaks of 61.332 keV, these x-ray fluorescence peaks and corresponding fluorescent escape peaks are also seen in the spectrum. 

\subsubsection{High-flux performance}
CdTe is known to polarize under high incident x-ray fluence.  Previous measurements on the CdTe first generation MM-PAD hybrid detector \cite{becker2016} have shown that polarization effects becomes apparent at 10$^{11}$ keV/mm$^{2}$ accumulated absorbed dose and are much more pronounced at 10$^{12}$ keV/mm$^{2}$ accumulated absorbed dose. The effects of polarization includes reduction in charge collection efficiency (count rate deficit), worsened image uniformity across the detector, and lateral movement of charge between polarized and unpolarized regions, resulting in degradation of the resolution and distortions. 

Polarization in the sensor material can be reset by letting the sensor sit under no radiation for an extended period of time, ranging from hours to days depending on the extent of polarization, or applying a forward bias for a short amount of time. Here, the sensor was reset by setting the sensor bias to -5 V for 1 minute, after which the bias was brought up to the operating value of the reverse bias at 400 V \cite{becker2016}. 

\begin{figure}
\centering 
\includegraphics[width = 0.7\textwidth]{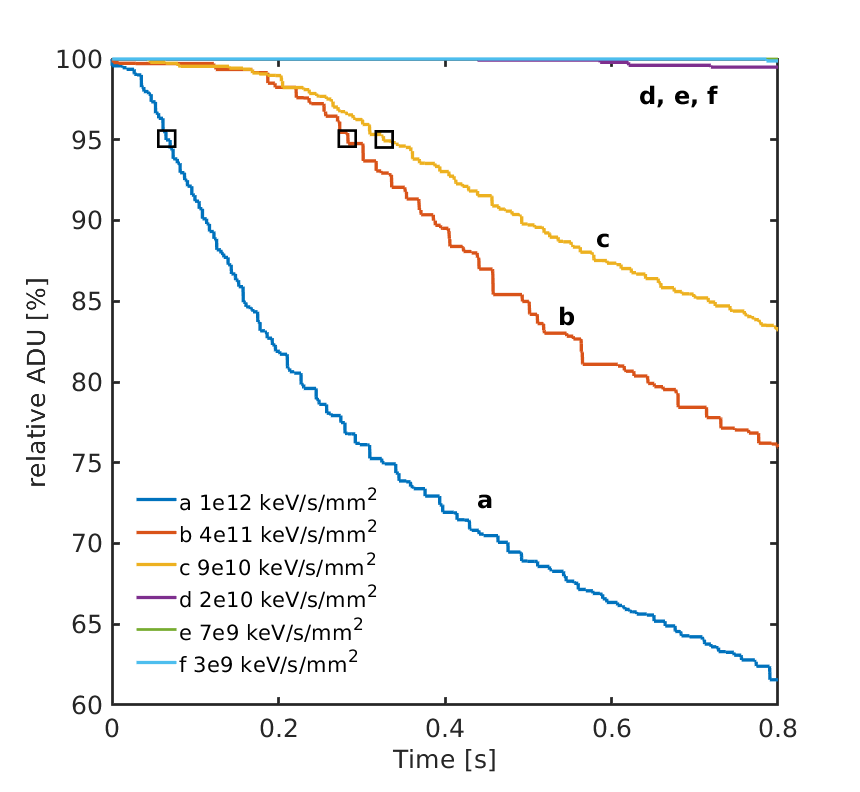}
\caption{\label{fig:attenscan} The normalized signal in the CdTe MM-PAD-2.1 decreases over time for high incident dose rates due to polarization. The black boxes indicate the time at which the signal reaches 95\% of its original value.}
\end{figure}

The effect of polarization, and in particular, the count rate deficit for different incident doses was measured for the CdTe MM-PAD-2.1. The incident flux was varied using six steel attenuators settings ranging from 0 mm (no attenuator) to 12 mm. Due to the presence of the second harmonic in the incident beam, the percentage of 122.664 keV photons increased in the incident beam as the attenuator thickness increased. The resultant incident flux varied from 10$^{12}$ keV/s/mm$^{2}$ to 3$\times$10$^{9}$ keV/s/mm$^{2}$ in a beam size set by slits to 600 $\upmu$m (height) by 2 mm (width). The spot size on the detector face was about 16$\times$6 pixel (FWHM) due to beam divergence. The frame integration time was set to 100 $\upmu$s and 10,000 frames were captured for each steel attenuator setting. Before illuminating the sensor and capturing these 10,000 frames for any attenuator setting, the sensor was first reset and de-polarized. Figure \ref{fig:attenscan} shows the normalized signal in the detector as a function of time, for various incident dose rates.  Polarization is observed to occur more quickly for higher incident dose. The effect of polarization can be measured in terms of `polarization time', which we define as the time it takes for the signal to drop to 95\% of its initial value. For an incident dose rate of 10$^{12}$ keV/s/mm$^{2}$, 4$\times$10$^{11}$ keV/s/mm$^{2}$ and 9$\times$10$^{10}$ keV/s/mm$^{2}$, the polarization time was measured to be 70 $\pm$ 20 ms, 290 $\pm$ 60 ms and 350 $\pm$ 120 ms respectively. For the dose rate of 2$\times$10$^{10}$ keV/s/mm$^{2}$ and below, no significant polarization was observed. This observation agrees with previous polarization studies done for CdTe \cite{becker2016}.

\section{Conclusions} 

Characterization measurements were conducted for the 128$\times$128 pixel Si and CdTe MM-PAD-2.1. The full-scale prototype of the MM-PAD-2.1 detector is planned to be a 2$\times$3 tiled system (256 $\times$ 384 pixels), each tile constituting a 128$\times$128 pixel hybrid module with CdTe as the sensor. The performance of the CdTe detector was tested for high energy x-rays under high and low flux at CHESS. Fluorescence in the CdTe was observed at low flux for an incident x-ray energy of 61.332 keV. For average beam energies lower than the fluorescence energies of Cd and Te the resolution performance of CdTe and Si MM-PAD-2.1 were comparable. The CdTe MM-PAD-2.1 detector showed polarization for an incident flux of $\sim$10$^{11}$ keV/s/mm$^{2}$ and higher, with the degradation in performance more rapid at higher incident flux. Future explorations include revision of the pixel layout by including radiation hard layout techniques, as well as the use of other high-Z sensors such as Ge which do not present the problem of polarization while providing high quantum efficiency even at high x-ray energies.

\acknowledgments
Support for detector development at Cornell has come from U.S. Dept. of Energy grants DE-FG02-10ER46693, DE-SC0016035, DE-SC0004079, and DE-SC0017631. This work is based upon research conducted at the Center for High Energy X-ray Sciences (CHEXS) which is supported by the National Science Foundation under award DMR-1829070. We are grateful to our  collaborators at the Advanced Photon Source (Argonne National Laboratory), especially Nino Miceli and John Weizeorick, for many useful discussions concerning this detector project.


\end{document}